\documentclass[]{aa}
\usepackage{psfig}
\usepackage{txfonts}
\begin{document}
   \title{Discovery of the heavily obscured supernova 2002cv}
   \author{A. Di Paola\inst{1}\and
           V. Larionov\inst{2,3}\and
           A. Arkharov\inst{4}\and
           F. Bernardi\inst{1,6}\and
           A. Caratti o Garatti\inst{1}\and
           M. Dolci\inst{5}\and
           E. Di Carlo\inst{5}\and
           G. Valentini\inst{5}
          }
   \offprints{Andrea Di Paola, \email{dipaola@mporzio.astro.it}}
   \institute{INAF - Osservatorio Astronomico di Roma (OAR),
    via Frascati 33, Monte Porzio Catone - Roma, Italy\and
    Astronomical Institute of St.Petersburg University,
    St.Petersburg, Petrodvorets, Universitetsky pr.~28, 198504, Russia\and
    Isaac Newton Institute of Chile, St.Petersburg Branch \and
    Central Astronomical Observatory, St. Petersburg, Russia \and
    INAF - Osservatorio Astronomico di Teramo (OACT), Teramo, Italy\and
    Universit\`a di Roma {\it Tor Vergata}, Roma, Italy
    }
   \date{Received 1 June 2002, Accepted 1 June 2002}

\abstract{On the 13th of May 2002, \object{supernova 2002cv} was
discovered using a near-infrared camera working at the AZT-24 1.1m
telescope at Campo Imperatore (AQ-Italy). After the infrared
detection a simultaneous photometric follow-up was started at
optical wavelengths. The preliminary results confirm a heavily
obscured object with a $V-K$ color not lower than 6 magnitudes,
making \object{SN 2002cv} the most reddened supernova ever
observed. This finding, along with the recent discovery of another
obscured supernova, suggests a critical revision of the rates
known to date. The estimate of the visual extinction and the light
curves are provided here. These latter indicate that our SN 2002cv
observations are the earliest available for a type-Ia supernova at
IR wavelengths.  } \maketitle

\section{Introduction}
\begin{figure*}[tbh]
 \centering
   \psfig{figure=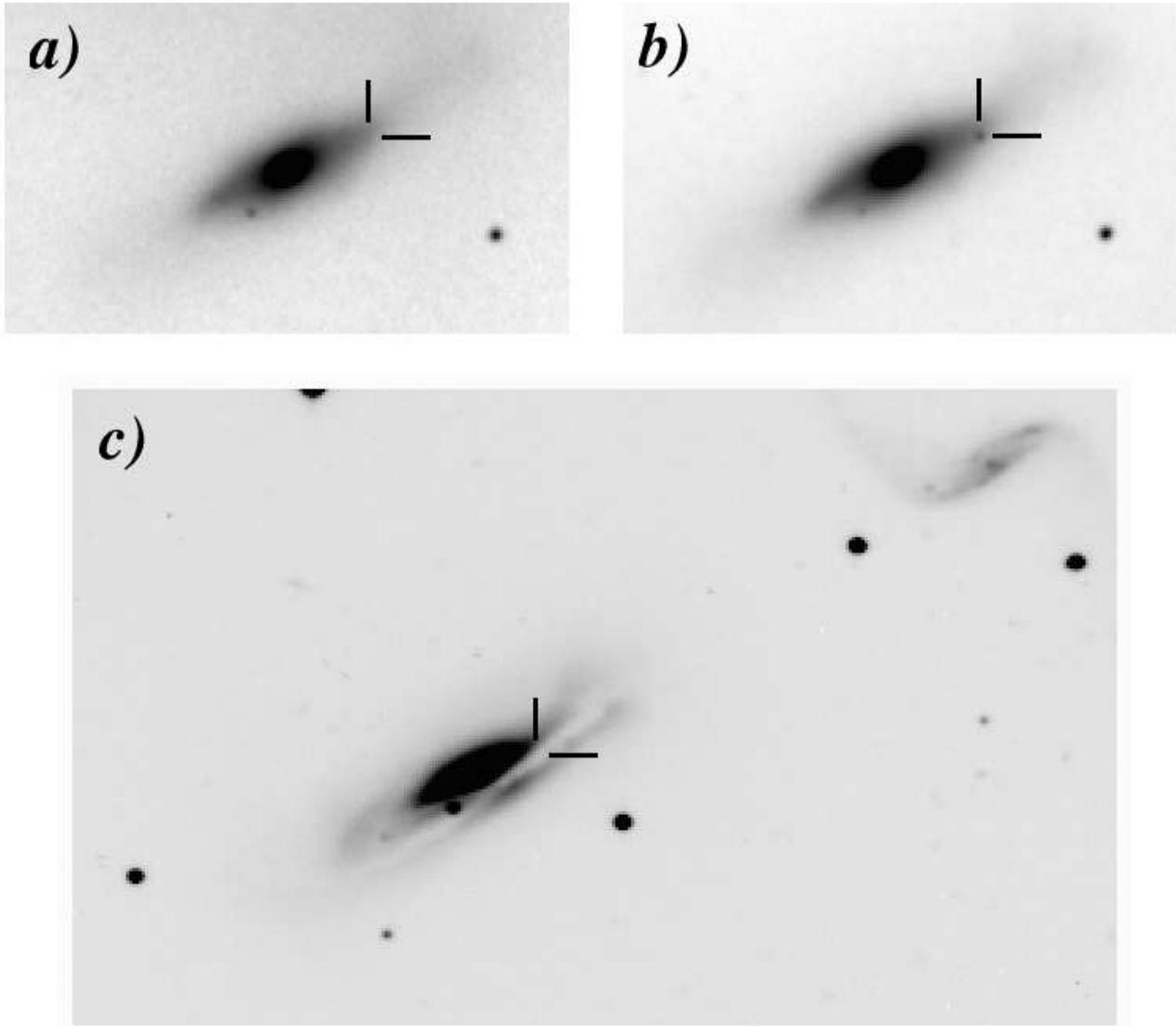, clip=}
   \caption{The $K$-band images obtained with AZT-24 telescope before
   (a) and after (b) \object{SN 2002cv} discovery. Panel (c) contains
   the $V$-band image obtained with the Schmidt telescope on May 15th
   (after the outburst) with no detectable object at the SN location.}
  \label{FigKV}
\end{figure*}

The AZT-24 telescope of the Campo Imperatore Observatory
\footnote{see {\it http://www.mporzio.astro.it
/cimperatore/WWW/}}
(a
 cooperation among Rome, Teramo and Pulkovo
 Observatories)
is mainly used for photometric studies of variable sources at near
infrared (NIR) wavelengths. During the follow-up of the
\object{supernova 2002bo} (Cacella et al. \cite{iauc7847}),
discovered on 2002 March 9 in the spiral galaxy \object{NGC3190},
a new source appeared $\sim28\arcsec$ to the North and
$\sim10\arcsec$ to the West of the galactic nucleus (Larionov et
al. \cite{iauc7901}) at $10^h18^m03\fs68$ and
$+21^\circ50\arcmin06\farcs2$ (J2000). Our discovery images of
this supernova, called \object{SN 2002cv}, are presented in
Fig.~\ref{FigKV}.

A more detailed analysis of the images obtained during the days
preceding May 13, that is the discovery date, has shown that the
outburst became visible between May 6 and May 9, though at
the limit of detectivity, thus preventing us from a prompt
detection of the new supernova. The maximum in our NIR light curves
was observed between the 20th and the 22nd of May ($J=14.77\pm 0.05$
on $MJD 52414.5\pm 0.5$, $K=13.92\pm 0.07$ on $MJD 52416.8\pm 0.5$).

As soon as the object was discovered, a simultaneous
follow-up was started at optical wavelengths using the Schmidt
telescope of the Campo Imperatore Observatory and the TNT
telescope of the Teramo Observatory.

Since our first discovery IAU Circular and the preparation of this text,
4 additional IAU Circulars were published about \object{SN 2002cv}.
In the first of them, Li (\cite{iauc7903}) reports the failed
localization of \object{SN 2002cv} using the KAIT
telescope\footnote{see {\it
http://astro.berkeley.edu/$\sim$bait/kait.html}} that observed
\object{NGC3190} at optical wavelengths ($B$, $V$, $R$ and $I$)
both before and after the discovery. In the second Meikle and
Mattila (\cite{iauc7911}) provide a preliminary classification of
the supernova as Type-Ia by means of the UKIRT telescope NIR
spectroscopy; this classification is confirmed in the fourth
Circular by Filippenko et al. (\cite{iauc7917}).

\section{Instrumentation}
The observations at NIR wavelengths were obtained by using the
1.1m telescope and the SWIRCAM camera (D'Alessio et al.
\cite{spie2000a}) at the Campo Imperatore Observatory (AQ-ITALY)
located 2150~m~a.s.l. SWIRCAM is based on a Rockwell PICNIC array
having $256\times 256$ pixels with a size of $40\mu \mathrm{m}$,
that corresponds to
 1.04\arcsec\ on the sky. Standard broad band filters ($J$,
$H$, $K$ and $Ks$) are available as well as some narrow band
filters and low resolution grisms ($R\approx 270$) for $IJ$ and
$HK$ spectroscopic observations.

The $60/90/180~\mathrm{cm}$ Schmidt telescope at Campo Imperatore
was used for optical follow-up. It is equipped with an optical CCD
camera (ROSI) using a $2048\times 2048$ pixels, back illuminated,
thinned array from Marconi Ltd. (Pedichini et al.
\cite{spie2000b}) featuring 1.5\arcsec\ per pixel on the sky.
Johnson broad band optical filters are available via a filter
jukebox especially designed for Schmidt telescopes focal station.

The TNT is an F/14 Ritchey-Chr\'etien $72~\mathrm{cm}$ telescope
located at the Teramo Observatory at about 300 meters a.s.l.,
equipped with Tektronic TK512 CB1-1 front illuminated $512\times
512$ pixels CCD camera and standard Johnson filters .

\section{Observations and Results}
\begin{figure*}
   \centering
   \psfig{figure=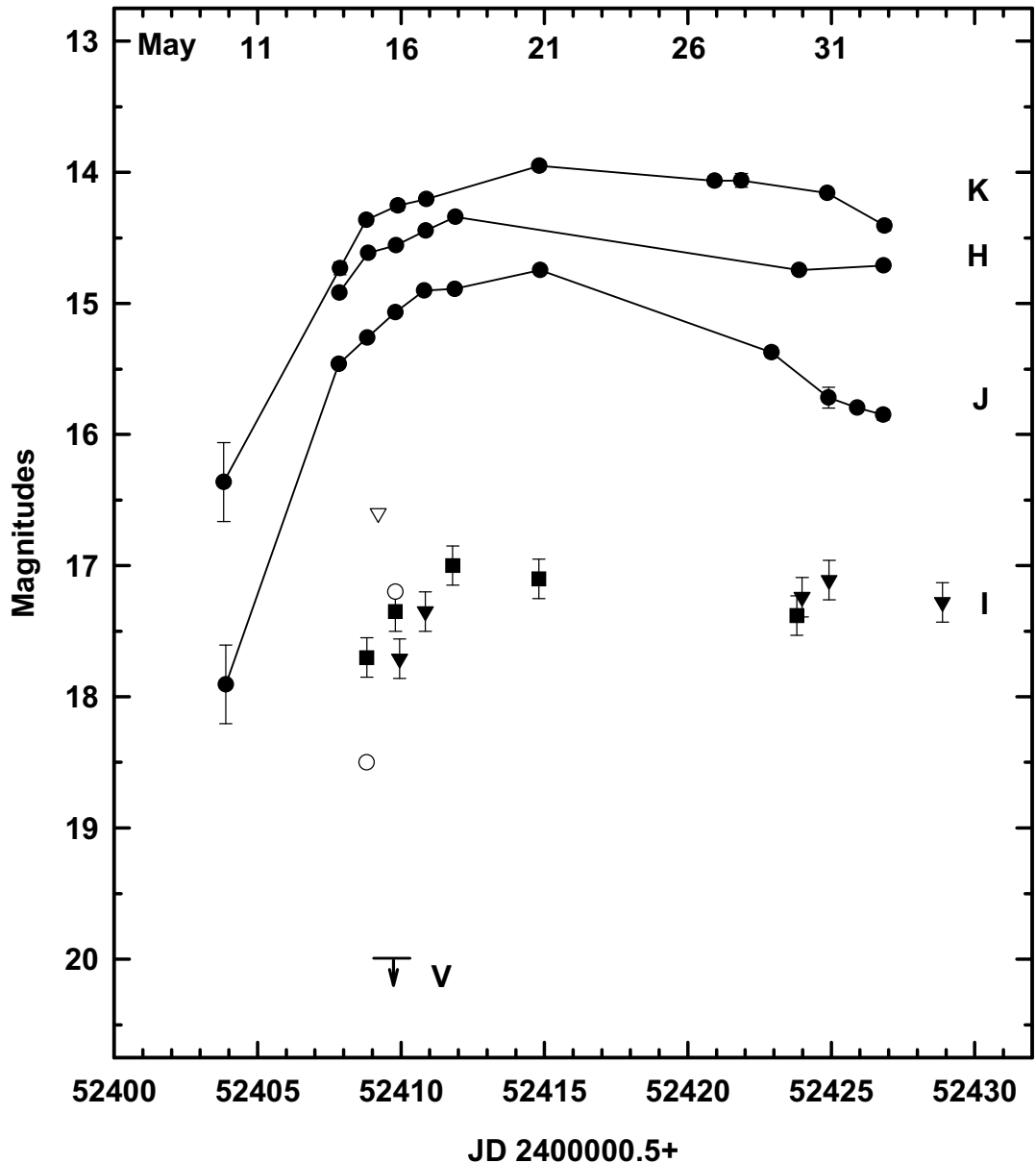, clip=}
   \caption{The preliminary light curves obtained during first 23 days
of \object{SN 2002cv} monitoring. Filled circles, connected
with straight lines, refer to AZT-24 NIR observations (one-sigma error
bars are shown; in most cases they are smaller than the symbol size).
Filled squares -- TNT $I$ band data, filled triangles -- Campo Imperatore
Schmidt telescope data, open triangle -- MMT data and open circles --
Nickel Telescope Lick Observatory data (the last two from Larionov et al.
\cite{iauc7901}). $V$-band upper limit, from Campo Imperatore
Schmidt telescope, is also shown.} \label{FigLC}
\end{figure*}

At NIR wavelengths, during the first 20 days of observations we
were able to acquire detailed light curves in $J$, $H$ and $K$
bands with very good signal-to-noise ratio (error bars are shown
on Fig.~\ref{FigLC}). At the same time $V$, $R$ and $I$ band
images were obtained both from the Schmidt and TNT telescopes.

It was immediately clear that \object{SN 2002cv} was heavily
obscured since it could not be detected in $V$-band images which
reach the 22nd magnitude on the sky and the 20th on the galaxy
disk (Fig.~\ref{FigKV}). This fact is supported by the existence
of a strong dust lane passing across the SN location which is
measured from IR
 images.

NIR images were obtained by 6 on-source ditherings with a
10\arcsec\ radius, as well as by side sky images. Both were
processed using the {\sl Preprocess} package (Di Paola
\cite{ppr}). The sky images were combined using 2-D median
composition to eliminate stars. The resulting sky frame was
subtracted from each target image before flat-fielding and
co-adding the dithered images. Because of the difficulties caused
by the galactic disk gradient just through the SN image, we
decided to perform plain aperture photometry providing galaxy
subtraction. Galaxy images without the SN are available from
images acquired before the \object{SN 2002cv} outburst, during the
\object{SN 2002bo} follow-up.
 The error estimates are derived from
the composition of the statistical errors on the supernova and
the standards measurements.

At optical wavelengths the Schmidt telescope images were also
dithered on a larger circle (5\arcmin). Only for $I$-band images
it was necessary to calculate a sky image on the basis of the same
target images, just to subtract the fringes pattern. In any case
all the images were flat-fielded and co-added before proceeding
with profile fitting photometry.

Unfortunately at optical wavelengths we have not a library image
as deep as those we obtained after \object{SN 2002cv} discovery.
This implies that the galaxy subtraction technique cannot be
applied and the accuracy of the photometry is affected by the
uncertainty in galaxy background contribution. This contribution
has been locally approximated by a second degree surface fitted on
the region surrounding the SN. The Landolt standards were used for
field photometric calibration.

Each TNT image was reduced by using standard bias subtraction and
flat-field normalization techniques. The photometric measurements
were performed by adopting a point-spread function (PSF) fitting
with the ROMAFOT package implemented in the MIDAS software. In
order to derive more accurate relative photometry of the supernova
with respect to two comparison stars located in the same field, we
have properly accounted for the host galaxy luminosity profile.
This was done during the fitting procedure by adopting a tilted
plane approximation for the diffuse background emission. The
calibration to the standard system was performed by using two
nearby standard stars from the Landolt catalogue.

All the available values are reported in the Fig.~\ref{FigLC}
 light curves (a further analisys will be published in a forthcoming paper).
From the plot it is clearly visible that \object{SN 2002cv} was
discovered while its brightness was still increasing, about 11
days before the IR maximum. According to Meikle and Mattila
(\cite{iauc7911}) and Filippenko et al. (\cite{iauc7917}) SN
2002cv is a Type-Ia. If this classification will be confirmed,
these observations will represent the earliest NIR measurements
available of such type of supernovae.

\section{Discussion}

It appears clearly from the near-infrared light curves and V-band
non-detection that \object{supernova 2002cv} is heavily obscured.
It is possible to evaluate the reddening basing on available
measurements and on the assumption it is a Type-Ia supernova. We
employ average optical--IR light curves from Nugent, Kim and
Perlmutter (\cite{NKP}) and standard galactic dust extinction law
(Rieke \& Lebofsky \cite{rieke}).

From Fig.~\ref{FigLC} we use MMT and Nickel Telescope data coupled
with our NIR simultaneous observations to obtain $I-K$ and $R-K$
colour indexes ($R\approx19^\mathrm{m}$ from MMT observations on
May 15). The resultant colour excesses correspond to values of
$A_\mathrm{V}$ in the range $7\fm4$ to $8\fm3$, while $V$ band
non-detection at $20^\mathrm{m}$ level sets a consistent lower
limit of $A_\mathrm{V}\geq7\fm0$. Another estimate of the
extinction may be derived from the SN absolute magnitude at
maximum. Since from our observations the $J$-band maximum is
better defined than others, we can obtain an estimate of
$A_\mathrm{J}$ using the SN apparent magnitude, the typical
absolute magnitude of Type Ia supernovae in the $J$-band from
Meikle (\cite{meikle}) and the distance modulus (DM) for the
galaxy. DM can be estimated both from \object{NGC3190} redshift
($z=0.00424$, see Heraudeau \& Simien \cite{heraudeau}) and from
Tully-Fisher method (see Tully \cite{tully}): the two values are
respectively $31.46\pm 0.2$ and $31.76$. In the following we will
then use $31.6\pm 0.3$. Then: $$A_\mathrm{J} = m_\mathrm{J} -
M_\mathrm{J} - DM = 14.8 + 19.0 - 31.6 = 2.2\pm 0.3.$$ The
accuracy depends mostly on the absolute magnitude and on the DM
uncertainty. Using $A_\mathrm{J}$, the $A_\mathrm{V}$ value is
obtained from the extinction law:

$$A_\mathrm{V} = A_\mathrm{J} / 0.282 = 7.8\pm 1.0$$

\noindent and confirms the previous estimates.

At a lower accuracy level, $A_\mathrm{V}$ can be estimated also
using $H$ and $K$ bands maxima. In these cases the results are
$A_\mathrm{V}=7.7\pm 2.3$ and $A_\mathrm{V}=10.0\pm 3.6$
respectively.

Definitely we estimate $A_\mathrm{V}=7.9\pm 0.9$, thus
\object{SN 2002cv} is really the most reddened supernova ever
observed.

If the complete light curves will support the hypothesis of
\object{SN 2002cv} to be overluminous supernova as suggested by
its affinity with \object{SN 1991T} (Li \cite{iauc7903},
Filippenko \cite{iauc7917}), then it strengthens the Howell
(\cite{howell}) correlation between the host galaxy morphological
type (\object{NGC3190} is classified as a SA(s)a late type galaxy)
and the supernova type-Ia luminosity.

We note that the observing of two supernovae simultaneously in the
same galaxy is supposed to be a very rare event according to the
actual supernova-rate estimates. This fact and the fact that this
object and \object{SN 2001db} (see Maiolino et al.
\cite{maiolino}) are invisible to most of the currently working
supernova search programs, seem to suggest that those
supernova-rates need to be significantly revised.

\section{Conclusions}

\begin{enumerate}
\item We present preliminary light curves of the \object{supernova
2002cv} we have discovered in \object{NGC3190} galaxy.
\item The supernova outburst appeared above magnitude $J=18$ between
the 6th and 9th of May, and reached its maximum on May 20, making
\object{2002cv} the earliest supernova Ia observed at NIR
wavelengths.
\item The color index $V-K\ga 6$ magnitudes makes this supernova
the most reddened ever observed.
\end{enumerate}

\begin{acknowledgements}
A.A. and V.L. are thankful to collegues from Roma and Teramo
Observatories for their kind hospitality during their observations
at Campo Imperatore. \end{acknowledgements}


\begin{thebibliography}{}

\bibitem[2002]{iauc7847}
Cacella, P., Hirose, Y., Chigasaki, et al., 2002, \iaucirc 7847

\bibitem[2000]{spie2000a}
D'Alessio, F., Di Cianno, A., Di Paola, A., et al., 2000,
\procspie, Volume 4008, 748

\bibitem[2000]{ppr}
Di Paola, A., 2000, Gamma-Ray Bursts in the Afterglow Era, 390, Roma 17-20 October 2000, Italy

\bibitem[2002]{iauc7917}
Filippenko, A.V., Chornock, R., Foley, R.J., \& Li, W., 2002,
\iaucirc 7917

\bibitem[1998]{heraudeau}
Heraudeau, P., \& Simien, F., 1998, \aaps, 133, 317

\bibitem[2001]{howell}
Howell, D.A., 2001, \apj, 554, L193

\bibitem[2002]{iauc7901}
Larionov, V., Arkharov, A., Caratti o Garatti, A., et al., 2002,
\iaucirc 7901

\bibitem[2002]{iauc7903}
Li, W., 2002, \iaucirc 7903

\bibitem[2002]{maiolino}
Maiolino, R., Vanzi, L., Mannucci, F., Cresci, G., Ghinassi, F.,
\& Della Valle, M., 2002, \aap, 389, 84

\bibitem[2000]{meikle}
Meikle, P., 2000, \mnras, 314, 782

\bibitem[2002]{iauc7911}
Meikle, P., \& Mattila, S., 2002, \iaucirc 7911

\bibitem[2002]{NKP}
Nugent, P., Kim, A., \& Perlmutter, S., 2002, \pasp, 114, 803

\bibitem[1985]{rieke}
Rieke, G. H., \& Lebofsky, M. J., \apj, 1985, 288, 618

\bibitem[2000]{spie2000b}
Pedichini, F., Speziali, R., D'Alessio, F., Di Paola, A., 2000,
\procspie, Volume 4008, 389

\bibitem[1988]{tully}
Tully, R.B., 1988, Cambridge University Press

\end{thebibliography}
\end{document}